# LUNAR IMPACT FLASH RESULTS AND SPACE SURVEILLANCE ACTIVITIES AT KRYONERI OBSERVATORY

**Alexios Liakos[1], Alceste Z. Bonanos[1], Emmanouil Xilouris[1], Detlef Koschny[2], Panayotis Boumis[1], Ioannis Bellas-Velidis[1], Richard Moissl[3], Athanassios Maroussis[1], Spyros Basilakos[1], [4], [5], Charalambos Kontoes[1]**

[1] *Institute for Astronomy, Astrophysics, Space Applications and Remote Sensing, National Observatory of Athens, Metaxa & Vas. Pavlou St., GR-15236, Penteli, Athens, Greece, Email: alliakos@noa.gr*

[2] *Chair of Astronautics, Technical University of Munich, 85748 Garching, Germany, Email: Detlef.koschny@tum.de*

[3] *Near Earth Object Coordination Centre, European Space Research Institute (ESA/ESRIN), Largo Galileo Galilei 1, 00044 Frascati (Roma), Italy, Email: richard.moissl@esa.int*

[4] *Academy of Athens, Research Center for Astronomy and Applied Mathematics, Soranou Efesiou 4, GR-11527, Athens, Greece, Email: sbasil@noa.gr*

[5] *School of Sciences, European University Cyprus, Diogenes Street, Engomi, 1516 Nicosia, Cyprus*

## ABSTRACT

We present current and future activities regarding lunar impact flash and NEO observations and satellite tracking from Kryoneri Observatory. In particular, we present results from the ESA-funded NELIOTA program, which has been monitoring the Moon for impact flashes since early 2017. Using the 1.2 m Kryoneri telescope, which is equipped with two high frame-rate cameras recording simultaneously in two optical bands, NELIOTA has recorded over 170 validated lunar impact flashes, while another ~90 have been characterized as suspected. We present statistical results concerning the sizes, the masses and the appearance frequency of the meteoroids in the vicinity of the Earth, as well as the temperatures developed during the impacts. Moreover, we present the capabilities of the Kryoneri telescope as a sensor for satellite tracking and the future plans regarding the provision of high-quality services for both the Planetary Defense activities of ESA (S2P/PDO) and the European Union's Space Surveillance and Tracking programme (EU/SST).

## 1 BRIEF DESCRIPTION OF THE KRYONERI OBSERVATORY

Kryoneri Observatory[1] (37°58′19″ N, 22°37′07″ E) is located at Mt. Kyllini, Corinthia, Greece at an altitude of 930 m (Fig. 1) and is operated by the Institute for Astronomy, Astrophysics, Space Applications and Remote Sensing[2] (IAASARS) of the National Observatory of Athens[3] (NOA). The observatory was established in 1972 and the telescope was installed in 1975 by the Grubb Parsons co. Ltd. In 2015, the telescope was upgraded (dome automatisation, electronic systems replacements, optical design transformation, new instruments installation) by DFM Engineering Inc. for the purposes of the NELIOTA project (see section 2).

The telescope is a prime focus reflector. Its primary mirror has a diameter of 1.2 m and a focal length of 3.6 m (f/3 focal ratio). The optical tube is on an off-axis german equatorial mount (Fig. 2). An instrument slider, which allows the interchange of the two current instruments setups, is installed at the prime focus of the telescope (Fig. 3).

The first observing setup consists of two twin front-illuminated sCMOS cameras (Andor Zyla 5.5) with a resolution of 2560×2160 pixels and a pixel size of 6.48 μm. The optical path is separated by a dichroic beam splitter into two wavelength regimes, below and above 730 nm. The cameras are placed at the end of each separated path. Each camera is equipped with one filter of the Johnson-Cousins photometric filter set. In particular, the red ($R_c$) filter, with a transmittance peak at $\lambda_R$=641 nm, is installed on the first camera, while the near-infrared ($I_c$) filter, with a transmittance peak at $\lambda_I$=798 nm, is set on the second camera. A focal reducer is installed in front of the dichroic reducing the effective focal length to 3.36 m (f/2.8 effective focal ratio). The field of view of this setup is approximately 17.0×14.4 arcmin$^2$. The second observing setup is an Apogee CG47 CCD camera (e2V CCD47-10 chip, 13 μm pixel size and a total of 1024×1024 pixels), equipped with one filter from the Bessell *UBVRI* photometric filter set. The field of view of this setup is approximately 12×12 arcmin$^2$.

An extensive description of the 1.2 m Kryoneri telescope and an evaluation of the instrument's performance can be found in [1]. The Kryoneri Observatory is planned to be totally upgraded with the addition of new instruments for the purposes of S2P/ESA and EU-SST programs. NOA aims to transform Kryoneri Observatory into a space hub in South-East Europe. More details are given in section 4.

---

[1] https://kryoneri.astro.noa.gr
[2] https://www.astro.noa.gr/
[3] https://www.noa.gr/

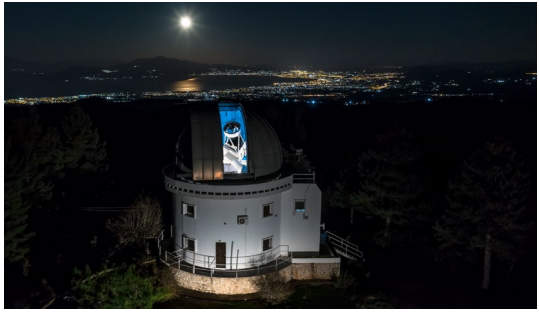

*Figure 1. Aerial view of Kryoneri Observatory.*

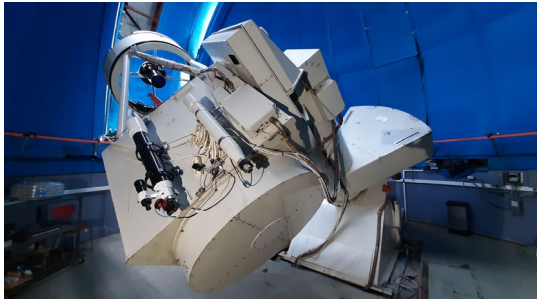

*Figure 2. The 1.2m Kryoneri telescope and its mount.*

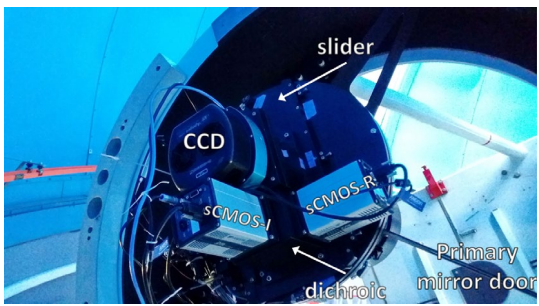

*Figure 3. The Prime Focus instrument of the Kryoneri telescope. The twin camera system consists of two identical sCMOS cameras fed via a dichroic beam splitter perpendicular to each other. The CCD camera is also shown. All cameras are attached onto a slider that allows the use either of the twin camera system or the CCD.*

## 2 LUNAR IMPACT FLASH CAMPAIGN AND RESULTS

### 2.1 Introduction

Near-Earth Objects (NEOs) are defined as asteroids or comets whose orbits come within 1.3 AU to the Sun. NEOs that intersect the orbit of the Earth may cause damage if they are large enough (typically above 10 m in size). These objects are listed e.g. on ESA's risk list[4], currently containing about 1400 objects. Meteoroids are small objects (i.e. less than 1 m in size) and they are remnants of asteroid collisions or cometary tails. These remnants orbit the Sun and form meteoroid streams, i.e. extended areas with a relatively high density of particles. When Earth enters these streams, meteor showers can be observed (e.g. Geminids, Perseids etc.). The randomly occurring meteors are called sporadics. Obviously, since the Moon follows Earth's orbit around the Sun, it is bombarded by the same stream or sporadic meteoroids.

The minor bodies in our solar system (i.e. asteroids, comets, meteoroids) orbit the Sun. Some of them may cross the orbit of the Earth and potentially pose a threat to our planet. However, the majority of the potential hazardous objects, i.e. large asteroids and comets (with sizes of the order of a few hundred meters up to several kilometres, so-called city destroyers and planet killers) have been detected and their trajectories are systematically updated (e.g. though the Minor Planet Center, NASA, ESA). The NEO risk list is frequently updated and the probability of such a body to hit Earth can be calculated. On the other hand, the small sizes of meteoroids do not allow such calculations, i.e. they reflect small amount of light to be easily observed. Obviously, the meteoroids cannot be considered as a serious threat for humans on Earth, given that the atmosphere plays the role of a shield. Except during the passage of the Earth through meteoroid streams, they also frequently appear as sporadics. Due to the increasing number of satellites and space missions, the collision of an artificial object with a meteoroid is not negligible anymore (e.g. the impact of micrometeoroids on JWST), especially during the passage of the Earth through the meteoroid streams. Moreover, the interest of space agencies for base establishments on other planets or moons (e.g. NASA-Artemis mission) has rapidly increased during the last decade. The Moon, and in the near future Mars, are the first targets. However, none of them has a dense atmosphere to protect the infrastructure and/or the astronauts from possible projectile impacts. Although meteoroids have small sizes (up to 1 m) and, thus, relatively low masses (up to a few tons), they travel with velocities from a few up to several tens of km s$^{-1}$. Therefore, their kinetic energies can reach the order of several billion Joules and can cause permanent and fatal damage to space-borne and ground-based equipment (on other planets) or even to the astronauts. However, although the sizes of the meteoroids vary, their majority have extremely small radii (up to a few cm), while their number increases inversely with their size. The latter means that small-size meteoroids appear more frequently than larger ones, as expected from the evolution of the solar system (i.e. the continuous collisions of asteroids with other minor bodies and comets perihelion passages form smaller and smaller meteoroids). The interest in their appearance frequency and size distribution has increased almost simultaneously with the increasing development of satellites and space missions, i.e. over the last two decades, due to the higher probability of a direct impact. [2] and [3] showed that there is a lack in knowledge for the flux densities of large meteoroids and small asteroids.

---

[4] https://neo.ssa.esa.int/risk-list

Meteoroids have been studied as meteors on Earth for many years. There is a lot of information about meteor showers in many websites that have been developed from both professional and amateur astronomers. These observations have provided us with valuable information regarding the origin of the meteoroids (i.e. parent body) of each shower and, in addition, their velocities. However, especially over the last two decades, studies of lunar impact flashes have increased because they can offer new information regarding the physical properties of the meteoroids.

The Moon can be considered as a perfect laboratory for studying meteoroid collisions. Due to the absence of an atmosphere, the meteoroids directly impact the lunar surface. Their kinetic energy is converted to a) heating the impacted material to temperatures where they emit light, b) kinetic energy of the ejecta, and c) crater excavation. The luminous energy, i.e. light emission, is responsible for generating the impact flash. The Moon can be observed using either ground-based or space telescopes orbiting either Earth or even the Moon [4]. In the case of observations from the Earth, a significant portion or even the whole lunar surface facing the Earth (approximately $19 \times 10^6$ km$^2$) can be recorded. Contrary to that, a meteor observer cannot cover more than a few thousands of square km on the sky, given that meteors are formed at an altitude of approximately 80-90 km. However, lunar observations have limitations due to the relatively low brightness of the flashes. Flashes on the lunar surface, caused by meteoroid impacts, can be observed only on the night-side part of the Moon because this region provides the necessary contrast (i.e. low background brightness). Depending on the size and the optical design of the telescope, the cameras and the filters used, this type of observation may occur, in general, between a lunar phase between 0.1 and 0.7. For higher illumination, the glare from the day-side part of the Moon is responsible for the increase of the background value of the night side, thus, the contrast decreases, which makes lunar impact flash detections impossible. Another significant factor that plays a crucial role to the impact flash detection is their short duration, i.e. from a few milliseconds up to a few seconds. Unfortunately, the majority last only few hundredths of a second, thus, fast-frame cameras are essential equipment for this type of observation.

To date, there are three groups performing systematic lunar monitoring for impact flashes. The first group is the MIDAS team, which operates in Spain ([5], [6], [7], [8]), the NASA's Meteoroid Environment Office [3], and the NELIOTA team. The first two teams utilize telescopes with mirror sizes between 30-50 cm and use no filters or single-band filters ([1], [5], [6], [7], [9]). However, the last few years, [10] and [11] have started to perform multi-wavelength observations.

## 2.2 NELIOTA project essential info

The *Near Earth objects Lunar Impacts and Optical TrAnsients* (NELIOTA) project began in early 2015 at the National Observatory of Athens (NOA) and has been funded by the European Space Agency (ESA). The first 1.5 years were spent for the telescope upgrade and the software development. This period was followed by a six-month testing period, while in February 2017 the observing campaign officially began. The first observing period lasted until January 2019 and was followed by a two-year extension of the program. The latter corresponds to the second observing period that lasted until January 2021. Between March and May 2021, the observations were unofficially continued. Finally, in June 2021, the NELIOTA team joined the *Consolidating Activities Regarding Moon, Earth and NEOs* (CARMEN) project, also funded by ESA, and the operations are planned to continue until June 2023.

The short-term goal of the project concerns the detection of lunar impact flashes, the temperature estimation of the impacts as well as the estimation of the physical parameters (mass, size) of the related meteoroids. The mid-to-long term goal is the determination the flux density of these objects in the vicinity of the Earth regarding their sizes and appearance frequency. The results will not only be scientifically useful, but they have a direct applicability for space agencies regarding the shielding of the satellites and space missions as well as for the future establishment of a base on the Moon. The NELIOTA team has published so far three papers in peer-reviewed journals [1], [12], [13] and a few more in conference proceedings [14], [15], [16]. Moreover, other research works are also based on the NELIOTA data [17].

NELIOTA operates the largest telescope worldwide dedicated to lunar impact flash detections. Using the 1.2 m Kryoneri telescope and two fast frame cameras (see section 1 and [1]), we are able to perform simultaneous observations in the red and infrared band-passes. The latter allows for the temperature estimation of the impacts and for the validation of the events using a single telescope ([12], [13]).

The observations are carried out between the 0.1 and 0.45 lunar illumination phases and point to the non-sunlit side part of the Moon. The upper limit is set by the glare produced from the sunlit side of the Moon that increases the background. Typically, the Moon can be observed between 4-8 nights per month split in two periods (i.e. before and after the new Moon), while the duration of the observations range between 30 min and approximately 5.5 h. During the lunar waxing phase, the telescope tracks the Moon up to an elevation of approximately 10°, which is the telescope's lowest pointing limit. On the contrary, given the peculiar mount of the telescope, i.e. the optical axis is eccentric with respect to the center of the dome, we cannot point to the Moon until its elevation is 20° during the lunar waning phase. The observations begin/end 20 min after/before the sunset/sunrise. Our setup covers approximately $3 \times 10^6$ km$^2$, but the exact coverage for each night depends on the Earth-Moon distance and the lunar phase. The observations are divided into 15 min chunks and before and after each chunk, a standard star is observed for magnitude calibration reasons (see also [12], [13], [16]).

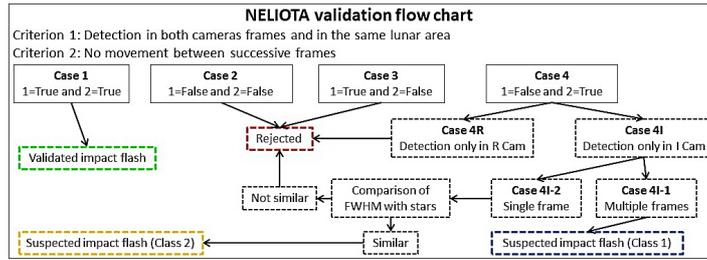

*Figure 4. NELIOTA validation flow chart. The chart includes all the possible cases of detections during the analysis process. The image was taken from [13].*

The cameras simultaneously record in the *R* and *I* bands at a frame rate of 30 frames-per-second (fps) in 2×2 binning mode (pixel scale ~0.8 arcsec pixel$^{-1}$). The exposures are 23 ms long and they are followed by a 10 ms pause time for the read-out. The cameras are synchronized with a typical accuracy better than 6 ms, while their absolute timing accuracy is less than 9 ms (see section 3).

For the lunar data acquisition and storage, a dedicated software, namely NELIOTA-OBS, has been developed ([1], [13], [16]). Briefly, this software calculates the observation plan of a night (i.e. lunar chunks, standard stars, calibration images), it creates metadata files for each chuck, and stores the data.

After the observations, another software, namely NELIOTA-DET, is used for the data calibration, reduction, and event detection. This software applies an algorithm that subtracts the lunar background from the images using a time-weighted average image of the preceding images (see [1], [13], and [16] for details). Thus, the frames that were taken exactly before a given image contribute more to the creation of the background image. After the subtraction, the software scans the residual image for sources that exceed a pre-set threshold. Once ten or more adjacent pixels are found to exceed this threshold, the software stores in another directory the current frame along with seven frames before and seven frames after the last detection.

Regardless on which camera's frames the detection was made, the software stores in the same directory the respective frames from the other camera too. This software also has a tool for the localization of the events. This tool is based on a cross match of the lunar features between the observed image and a highly detailed map of the Moon ([12], [13], [16]).

The validation of the events is made by an expert user. The user inspects all the detected events one by one and based on the NELIOTA's validation flow chart (Fig. 4) judges whether an event is: a) validated, b) suspected, or c) junk (e.g. cosmic rays, satellites). More details for the validation process and for ruling-out possible slow-moving satellite glints can be found in [13] and [16].

For the validated and suspected flashes, we apply aperture photometry to calculate their photon fluxes. These fluxes are converted into magnitudes using the standard stars observations between the chunks ([12], [13]).

The data for the detected lunar impact flash detections are uploaded within 24 hours of observations into the project's official webpage[5] using the NELIOTA-ARC software. The data are publicly available, thus every visitor can download the images of the validated flashes. However, registered users can access the data of the suspected flashes too. The data can be used either for scientific or educational purposes.

The NELIOTA team, using its long-term experience on lunar impact flash detections, developed and in December 2022 publically released the *Flash Detection Software*[6] (FDS). This software was designed for amateur astronomers and can be used with a variety of equipment (cameras and telescopes) to detect impact flashes. It aims to increase the number of lunar impact flash observers worldwide in order to increase the number of observed flashes and, thus, to help us obtain a more complete statistics. The software can be also used for flashes on planets (e.g. Jupiter, Mars). The next steps regarding the validation of heterogeneous data and their archiving are discussed in section 4.

---

[5] https://neliota.astro.noa.gr/
[6] https://kryoneri.astro.noa.gr/en/flash-detection-software/

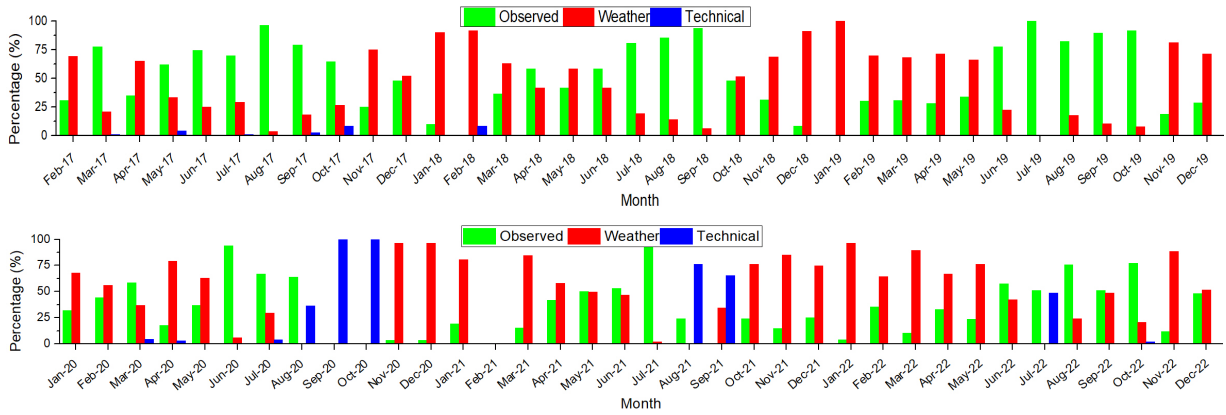

*Figure 5. Observing efficiency of the NELIOTA campaign since February 2017. The upper panel shows the statistics per month (the percentages are based on the total available time for lunar observations) between February 2017 and December 2019. The lower panel includes the respective statistics but for the range January 2020 – December 2022.*

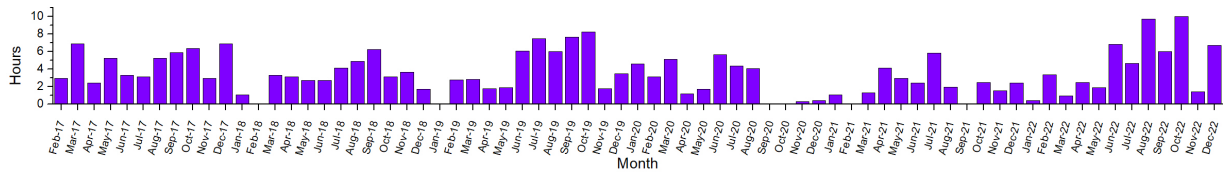

*Figure 6. Total actual observed hours of lunar monitoring per month since the beginning of the NELIOTA campaign. The bars of this plot correspond to the respective green bars of Fig. 5.*

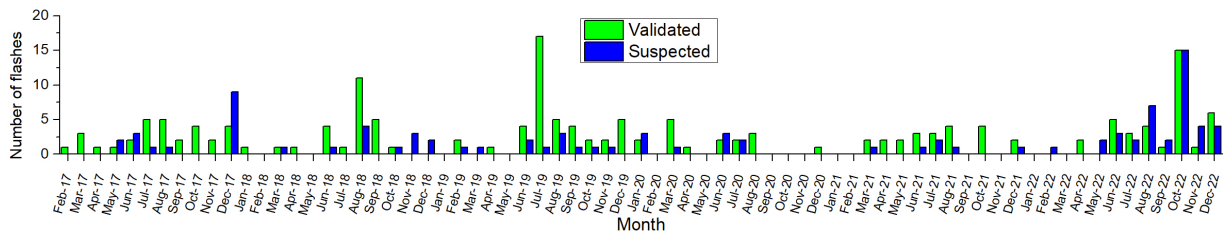

*Figure 7. Lunar impact flashes detected from the NELIOTA campaign from February 2017 up to December 2022. The green bars correspond to the validated and the blue bars to the suspected flashes (for method details see Fig. 4).*

## 2.3 Operations statistics and efficiency of the campaign

The NELIOTA observations began in February 2017 and continue to date. Fig. 5 shows the statistics of the campaign regarding the percentages of the available time for lunar observations that were: a) actual observations, b) lost due to bad weather conditions, and c) lost due to technical reasons. Fig. 6 shows the distribution of observation hours over time since the beginning of the campaign. It should be noted that, in this plot, the observation hours per month correspond to the true recording time of the Moon and not to the total available time. Due to the cameras' read-out time, 30% of the total available time is lost. Moreover, approximately another 8% is lost due to the repositioning of the telescope every 15 min for the standard stars observations. Therefore, a total of approximately 38% of the total available time is lost. Fig. 7 shows the statistics of detections during the campaign including both the validated and suspected flashes. Fig. 8 shows the pie chart of the efficiency of the NELIOTA campaign. It includes both the absolute values of hours and the corresponding percentages of the actual recording time, the time lost due to bad weather conditions, and the time lost due to technical problems. It should be noted, that this chart concerns only the time dedicated to lunar observations and not the global weather statistics of Kryoneri Observatory.

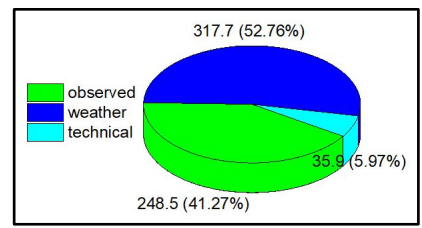

*Figure 8. Pie chart of the efficiency of NELIOTA campaign between February 2017 - December 2022. The chart concerns only the time of lunar observations.*

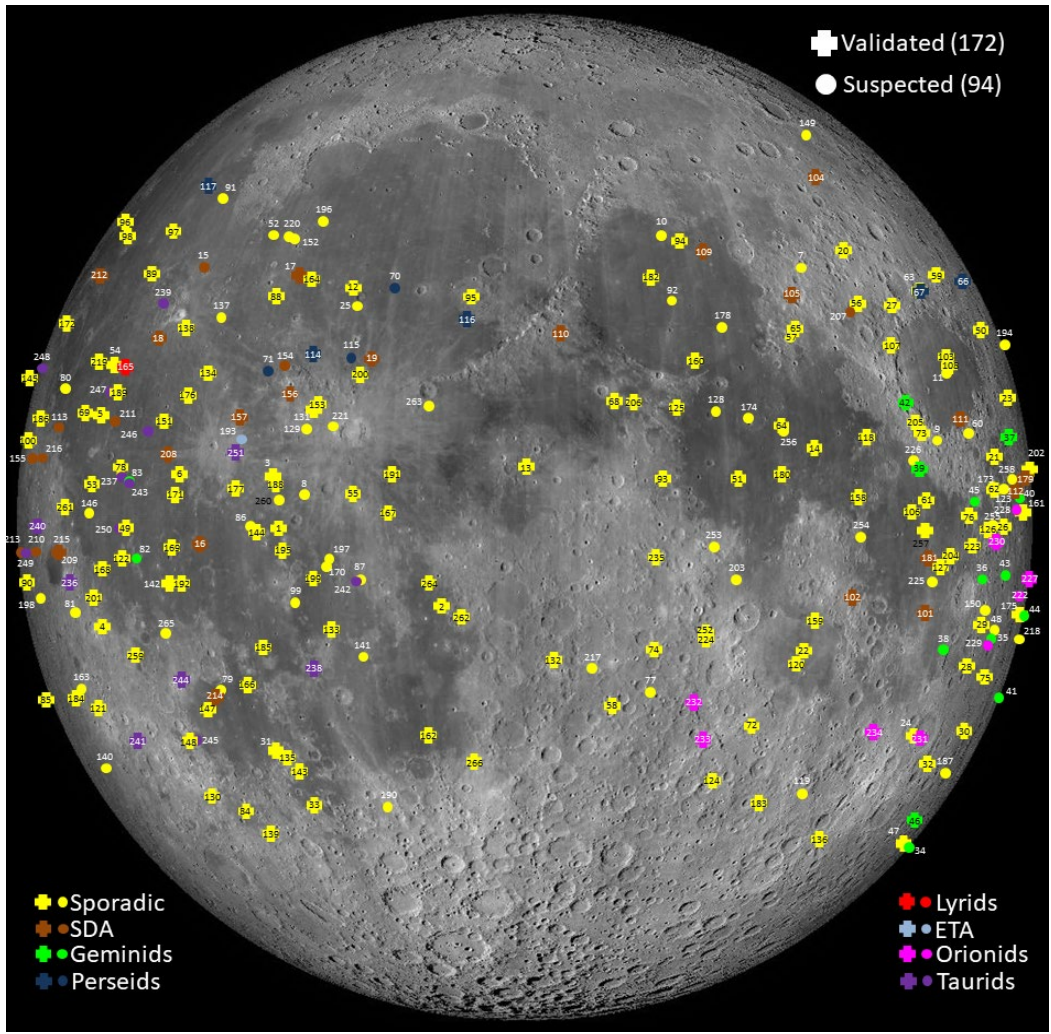

*Figure 9. Positions of the detected flashes by the NELIOTA campaign between February 2017-December 2022. The crosses denote the validated flashes, while the filled circles the suspected ones. Yellow symbols correspond to the sporadic flashes. Colors other than yellow denote the possible association of the projectiles with an active meteoroid stream during the time of the observations. The names of the streams are given on the lower left and right corners. The numbering of the flashes is in chronological order starting from #1 (01 February 2017, 17:13:57.9 UT) and they are in accordance with the data repository of the NELIOTA website (for registered users).*

Finally, Fig. 9 illustrates the positions of all the detected flashes on a highly detailed lunar map (for localization method see section 2.2 and [12], [13], [16]) taken from the *Virtual Moon Atlas software*[7]. The colors of the symbols denote whether the impacting meteoroid has been associated with an active meteoroid stream (for the methodology see [7], [8], [13], [18]).

## 2.4 Results

NELIOTA observations aim to derive: a) the physical properties of the projectiles (mass and size), b) the temperatures of the impacts (and their evolution in time), and c) the expected crater sizes. Moreover, we aim to obtain the statistics of these objects regarding their appearance frequency in the vicinity of the Earth.

The kinetic energies ($E_{kin}$) of the meteoroids are calculated with a standard formalism described in many papers ([2], [3], [5]-[13] and references therein). Briefly, we use the magnitudes of the flashes to calculate their luminous energy ($E_{lum}$). Given that NELIOTA operates in dual band observing mode, we calculate the $E_{lum}$ for each band. Moreover, according to the method described by [13, sect. 6.2.2] regarding the energy correction (i.e. calculation of the energy not measured due to the read-out time of the cameras in the cases of multi-frame flashes), the energy of each band, as determined from the photons fluxes, and the respective calculated (i.e. lost) energies are added together in order to determine the total luminous energy emitted from the flash. Subsequently, using the definitions of [5], [7], [9], and [10] regarding the luminous efficiency $\eta$ (where $\eta=E_{lum}/E_{kin}$), and after applying it to our dual band observations (see [13, section 6.2.1]), we are able to calculate the kinetic energies of the projectiles assuming a typical value of $\eta=1.5 \times 10^{-3}$ ([9], [19]).

---

[7] https://www.ap-i.net/avl/en/start

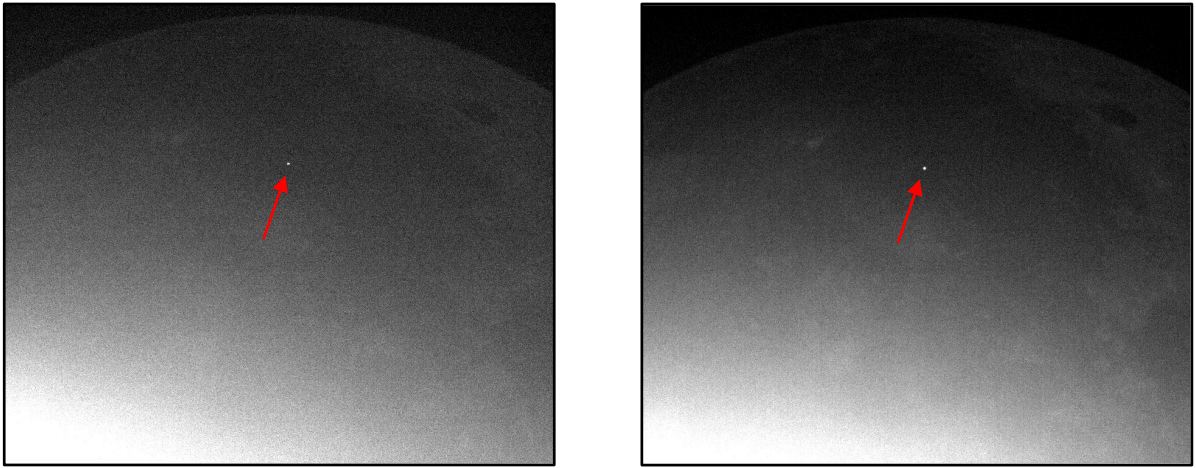

*Figure 10. Sample of a validated lunar impact flash (#151) detected on 25-Jun-2020 18:28:18.3 UT during the NELIOTA operations. The left image is from the Red-filtered camera and the right image from the Infrared-filtered camera. This flash produced peak magnitudes $m_R$=7.92 mag and $m_I$=6.66 mag in R and I bands, respectively, a peak temperature of 2760 K and its maximum duration was 132 ms.*

The velocities ($V$) and the bulk densities ($\rho$) of the meteoroids are assumed according to their origin (for the meteoroid stream association method see [7], [8], [18], and also [13] for a brief description). For the sporadic ones, a velocity value between 17-24 km s$^{-1}$ and a density value of 1.8 g cm$^{-3}$ are assumed ([2], [3], [5]-[13] and references therein). For meteoroids that have been associated with streams, we use the proposed values from the literature (c.f. [13, Table E.1]). Thus, the masses ($m$) of the impactors are derived using the standard definition of kinetic energy (i.e. $E_{kin}=1/2\ mV^2$), while their radii ($r$) according to the density-mass-volume relation (i.e. $r=(3m/4\pi\rho)^{1/3}$ for a sphere). Finally, the crater sizes are calculated using the scaling law of [20] (c.f. [21], [13]) assuming an average impacting angle of 45°. The temperatures ($T$) of the impacts are calculated using the Planck's black body law and the ratio of the photons flux per band, e.g. see [12], [13]. Temperatures can be calculated only for the validated flashes (i.e. those observed in both cameras' frames). Moreover, for multi-frame flashes that have been recorded in both cameras for more than one set of frames, it is feasible to calculate the temperature evolution of the impact, i.e. its cooling curve.

Results for the first 79 validated and 33 suspected flashes (up to July 2019) detected by the NELIOTA campaign can be found in [13]. The respective detailed results for the rest will be published in a near-future work of the NELIOTA team. However, we present herein a small randomly selected sample of ten consecutively detected validated flashes after July 2019. Fig. 10 shows the detection of a validated flash in both cameras' frames, while Fig. 11 illustrates its light curves and its cooling curve. Tab. 1 includes the observational results for ten validated flashes, i.e. their timings of occurrence, duration (dt), peak magnitudes, total emitted

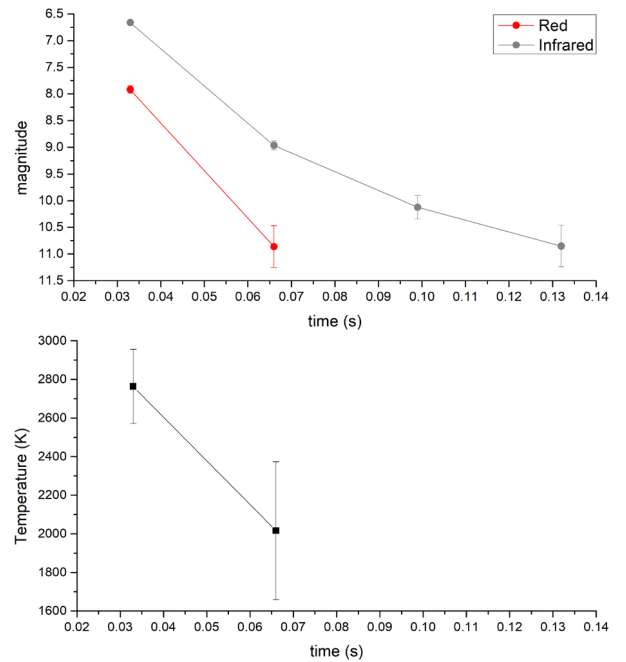

*Figure 11. Upper panel: Light curves of the lunar impact flash #151 (Fig. 10) in R and I bands. Lower panel: The cooling curve of this flash, which corresponds to the first two sets of R and I frames. The temperature dropped approximately 750 K in 33 ms.*

---

*Table 1. Observational results for a sample of ten validated flashes detected by the NELIOTA campaign between March and July 2020. For the methods followed for the derivation of the parameters and for the errors calculations and estimations see [13]. For the numbering of the flashes see legend of Fig. 9. The term "spo" denotes the sporadic origin of the meteoroids and the "SDA" the association with the Southern Delta Aquariids meteoroid stream.*

| # | Date & time (UT) | $dt$ (ms) | $m_R$ (mag) | $m_I$ (mag) | $E_{lum}$ ($\times 10^4$ J) | Origin | Latitude (deg) | Longitude (deg) |
|---|---|---|---|---|---|---|---|---|
| 142 | 2020 03 01 16:54:23.9 | 99 | 8.32±0.08 | 7.15±0.04 | 12.7±0.4 | spo | -7.8±0.5 | -45.1±0.5 |
| 143 | 2020 03 01 17:10:06.4 | 33 | 9.92±0.25 | 9.42±0.12 | 1.5±0.2 | spo | -30.6±0.5 | -31.3±0.5 |
| 144 | 2020 03 27 17:40:25.3 | 66 | 10.01±0.15 | 8.70±0.08 | 2.5±0.1 | spo | -1.9±0.5 | -32.1±0.5 |
| 145 | 2020 03 29 18:14:10.9 | 66 | 10.83±0.23 | 9.72±0.09 | 1.0±0.1 | spo | 15.1±0.5 | -91.3±0.5 |
| 147 | 2020 03 29 19:16:46.5 | 33 | 10.18±0.21 | 9.26±0.09 | 1.4±0.1 | spo | -22.7±0.5 | -42.8±0.5 |
| 148 | 2020 04 28 19:19:54.5 | 66 | 8.99±0.10 | 8.13±0.05 | 5.5±0.2 | spo | -26.6±0.5 | -47.7±0.5 |
| 151 | 2020 06 25 18:28:18.3 | 132 | 7.92±0.10 | 6.66±0.06 | 18.1±0.6 | spo | 10.8±0.5 | -46.5±0.5 |
| 153 | 2020 06 26 19:52:31.5 | 66 | 9.73±0.11 | 8.22±0.07 | 3.5±0.1 | spo | 12.7±0.5 | -24.9±0.5 |
| 156 | 2020 07 26 19:08:21.2 | 33 | 10.20±0.12 | 9.13±0.12 | 1.3±0.1 | SDA | 14±0.5 | -28.6±0.5 |
| 157 | 2020 07 26 19:10:25.4 | 66 | 9.15±0.10 | 7.82±0.06 | 5.2±0.2 | SDA | 11.3±0.5 | -35.2±0.5 |

*Table 2. Physical parameters of the meteoroids that produced the lunar impact flashes listed in Tab. 1, peak temperatures of the flashes and expected crater diameters.*

| # | $E_{kin}$ ($\times 10^7$ J) | $V^*$ (km s$^{-1}$) | $\rho^*$ (g cm$^{-3}$) | $m$ (g) | $r$ (cm) | $T$ (K) | $d_c$ (m) |
|---|---|---|---|---|---|---|---|
| 142 | 8.4±0.2 | 17 | 1.8 | 585±16 | 4.2±1.4 | 2919±161 | 3.6±0.1 |
| 143 | 1.0±0.1 | 17 | 1.8 | 70±10 | 2.1±7 | 4991±2002 | 1.9±0.1 |
| 144 | 1.7±0.1 | 17 | 1.8 | 117±7 | 2.5±8 | 2677±263 | 2.2±0.1 |
| 145 | 0.7±0.1 | 17 | 1.8 | 48±4 | 1.8±6 | 3030±507 | 1.7±0.1 |
| 147 | 0.9±0.1 | 17 | 1.8 | 65±6 | 2.0±7 | 3430±605 | 1.9±0.1 |
| 148 | 3.7±0.1 | 17 | 1.8 | 255±10 | 3.2±1.0 | 3587±323 | 2.8±0.1 |
| 151 | 12.1±0.4 | 17 | 1.8 | 836±29 | 4.7±1.6 | 2763±192 | 3.9±0.1 |
| 153 | 2.3±0.1 | 17 | 1.8 | 161±7 | 2.7±9 | 2408±160 | 2.4±0.1 |
| 156 | 0.9±0.1 | 41 | 2.4 | 10±1 | 1.0±9 | 2788±365 | 1.9±0.1 |
| 157 | 3.5±0.1 | 41 | 2.4 | 41±2 | 1.6±8 | 2649±171 | 2.9±0.1 |

*assumed

luminous energy ($E_{lum}$), most possible origin, and location on Moon (Latitude and Longitude). Tab. 2 lists parameters of these flashes based on the assumptions described in the previous paragraphs. Particularly, this table includes the physical parameters of the impacting meteoroids ($m$, $r$, $E_{kin}$), their assumed velocities ($V$) and densities ($\rho$), the peak temperature values of the impacts ($T$) and the resulting expected crater diameters ($d_c$).

Fig. 12 illustrates the magnitude distributions of all the flashes detected by the NELIOTA campaign up to December 2022, while Fig. 13 shows the statistics of the physical parameters of only the meteoroids that produced validated flashes. The two plots of Fig. 13 show that the majority of the meteoroids have masses less than 200 g and radii less than 4 cm. It should to be noted that ~63% of this sample have masses less than 100 g. The statistics for the impact peak temperatures and the expected crater sizes are given in Fig. 14. The majority (~85%) of the new craters have diameters between 1.5-3.5 m. The temperatures range between 1400-7450 K, with 61.5% to lie inside the 2000-3500 K regime, and 85.5% between 2000-4500 K.

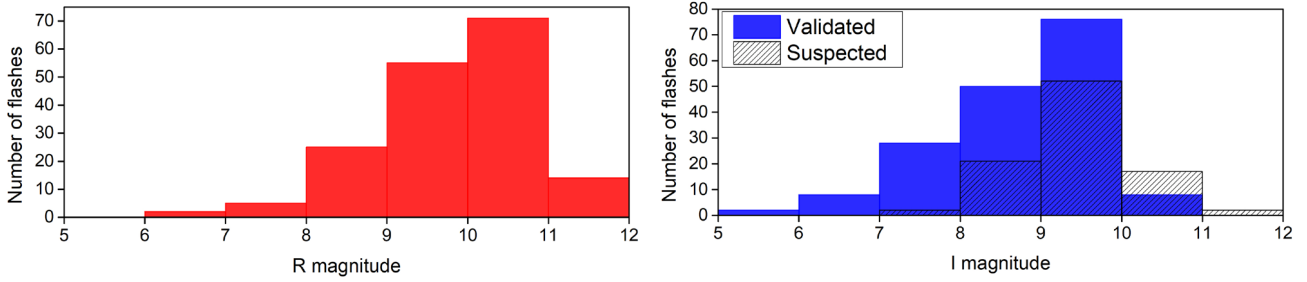

*Figure 12. Magnitude distributions of the detected flashes. The left image corresponds to the R filter, while the right image to the I filter. The respective (stand-alone) distribution of the suspected flashes is also plotted in the right image. According to Fig. 4 there are not any suspected flashes detected only in the R pass band (see also [13]).*

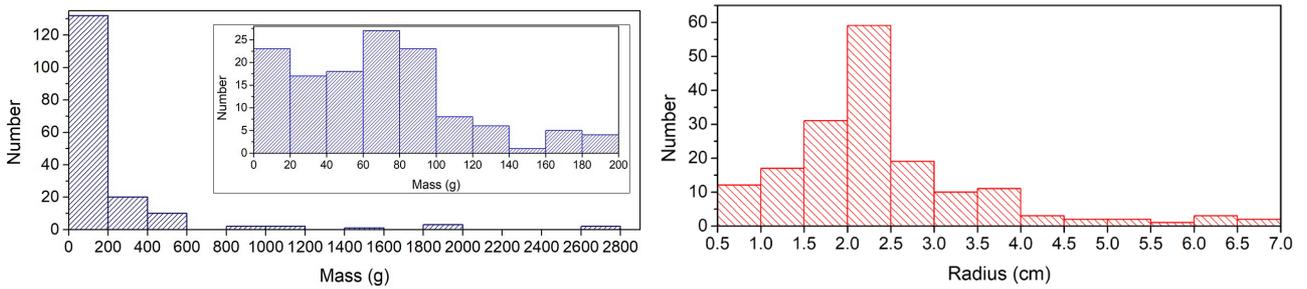

*Figure 13. Statistics of the physical parameters of the meteoroids that produced validated lunar impact flashes and were detected by the NELIOTA campaign (172 in total). The left image shows the distribution of the masses and includes also an internal panel with the respective distribution for masses up to 200 g. The right image shows the distribution of the radii. For the assumptions regarding the velocities, densities, and luminous efficiency see text.*

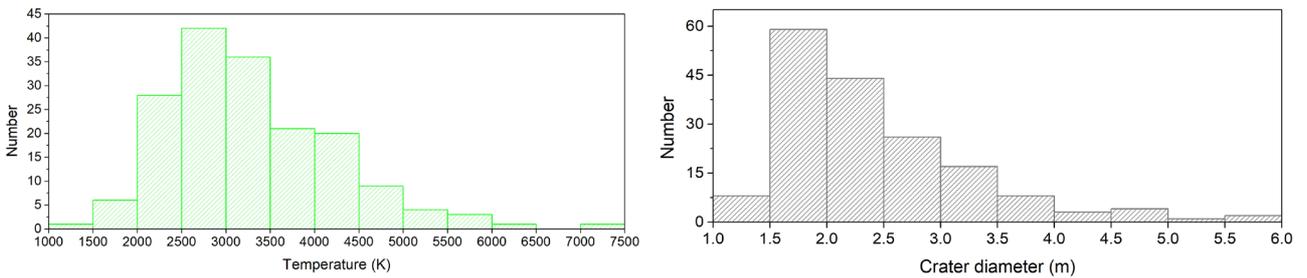

*Figure 14. Left image: Distribution of peak temperatures evolved during the meteoroid impacts on Moon that were detected by the NELIOTA campaign. Right image: Distribution of the estimated crater sizes as results of the impacts assuming an impacting angle of 45°.*

By the end of December 2022, NELIOTA has observed the Moon for a total of 248.5 h (see Fig. 8) and has detected 172 validated lunar impact flashes. From this observing time, 36.5 h occurred when the Earth-Moon system was inside strong meteoroid streams, i.e. those with high zenithal hourly rates, such as Geminids, SDA, Perseids, Orionids (see [13, Table E.1]). According to the meteoroid stream association method of [7], [8], and [18] (see also [13] for a brief description), 43 flashes have been identified to originate from projectiles of active meteoroid streams (see Fig. 7 for increased flash detections in particular months). Based on these results, we are able to calculate the lunar impact flash detection rates of NELIOTA. Furthermore, using these rates, the respective appearance frequencies of meteoroids on the Moon and in the vicinity of the Earth can be also calculated (see [13] for method). Tab. 3 lists the aforementioned rates and appearance frequencies estimations for the total surface of the Moon and for various distances from the Earth's surface for the cases when the Earth-Moon system is or is not inside a strong meteoroid stream. The distances from the Earth's surface correspond to the mesosphere (90 km), to the upper limit (1600 km) of the low-Earth orbit (LEO) zone, to the middle of medium-Earth orbit (MEO) zone (20000 km) and to the geosynchronous orbit (GEO; 36000 km). The meteoroid appearance frequency in the vicinity of the Earth is plotted in Fig. 15.

Table 3. NELIOTA detection rates, based only on the validated flashes, and meteoroid appearance frequencies on Moon (total surface) and around Earth when the Earth-Moon system is not (sporadic) or is inside a meteoroid stream.

|  | Detection rate (meteoroid h$^{-1}$ km$^{-2}$) | Impact frequency on Moon (meteoroid h$^{-1}$) | Appearance frequency around Earth (meteoroid h$^{-1}$) | | | |
| --- | --- | --- | --- | --- | --- | --- |
|  |  |  | 90 km | 1600 km | 20000 km | 36000 km |
| Sporadic | 1.93×10$^{-7}$ | 7.3 | 101 | 154 | 1683 | 4344 |
| Stream | 3.79×10$^{-7}$ | 14.4 | 199 | 303 | 3312 | 8550 |

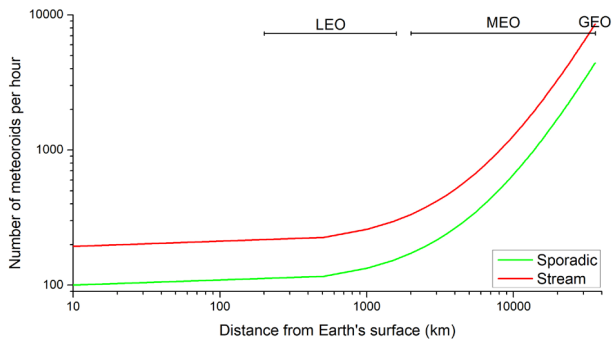

Figure 15. Appearance frequency of meteoroids around Earth when Earth-Moon system is (red line) or is not (green line) inside a meteoroid stream. Low-Earth orbit (LEO), medium-Earth orbit (MEO), and geostationary orbit (GEO) ranges are also indicated.

The NELIOTA detection rates can be calculated by simply dividing the total number of the detected flashes with the observed hours and the mean surface coverage on Moon. Taking into account only the validated flashes (sporadic and stream associated), the detection rate is derived as 2.23×10$^{-7}$ meteoroid h$^{-1}$ km$^{-2}$, while by including the suspected flashes too, this rate can reach the value of 3.44×10$^{-7}$ meteoroid h$^{-1}$ km$^{-2}$. Omitting the covered area, the NELIOTA detection rates (based only on validated flashes) can be calculated as 0.6 and 1.18 meteoroid h$^{-1}$ for sporadic and stream meteoroids, respectively. It should to be noted that the magnitude histograms (Fig. 12) indicate an increase in numbers at fainter magnitudes, i.e. we have not yet reached the peak.

A comparison with similar campaigns ([3], [22], i.e. between 1.03-1.09 meteoroid h$^{-1}$ km$^{-2}$) shows that NELIOTA has at least doubled the detection rate of lunar impact flashes, while is able to detect approximately two magnitudes fainter flashes, which corresponds approximately to one order of magnitude less in mass values. Obviously, the reason for this increase in the detection rate is the larger diameter telescope used and the systematic monitoring. Therefore, NELIOTA can be considered as the most efficient campaign for lunar impact flashes worldwide.

## 3 THE 1.2m KRYONERI TELESCOPE AS A SATTELITE TRACKING SENSOR

Greece entered the pre-accession negotiations for membership in the EU-SST partnership in late 2020. The National Observatory of Athens (NOA) led the greek side and supported the full membership status of Greece in the partnership.

The 1.2 m Kryoneri telescope participated in the third calibration campaign (CCW#3) of the EU-SST with a cut-off date by the end of January 2021. The requested precision of the satellites positions was less than 3.6 arcsec. The ESA evaluators concluded that the number of the provided measurements and the positions precision lie well inside the requirements of EU-SST tracking operations. Particularly, our measurements had an astrometric precision of 0.5 arcsec and a time offset less than 9 ms. Therefore, the 1.2 m Kryoneri telescope was accepted as an EU-SST sensor for MEO and GEO satellite monitoring in tracking mode. NOA has been as-signed as the national entity/representative of Greece in the EU-SST partnership and signed a memorandum of understanding (MoU) in April 2022. Since September 2022, Greece is a full member of the EU-SST partnership.

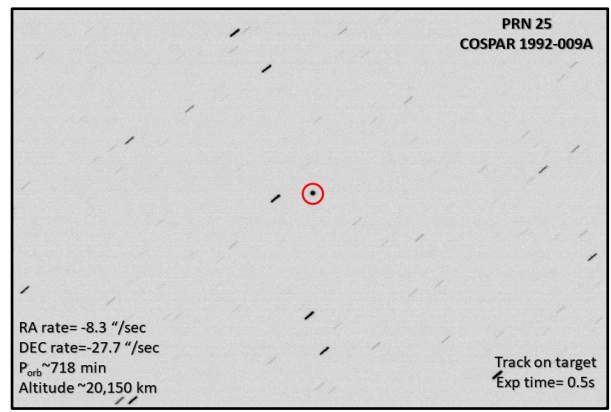

Figure 16. Sample image of testing observations of the MEO satellite PRN 25 using the 1.2 m Kryoneri telescope and the I-filtered sCMOS camera. The satellite is the point source inside the red circle, while the field stars appear as streaks. The observational details are also given in the lower corners of the image.

Currently, the Greek team working on the EU-SST programme is being established and is rapidly increasing with the addition of new members. Except for the 1.2 m telescope operations, the team will also work on the Research and Development (R&D) projects of the EU-SST. Currently, there are ongoing tasks at the Kryoneri Observatory regarding the forthcoming EU-SST operations. These preparations, along with the preliminary tests on real targets, are scheduled to last until June 2023. Finally, the 1.2 m telescope is scheduled to be fully operational for the EU-SST purposes in July 2023.

## 4 FUTURE ACTIVITIES AT KRYONERI OBSERVATORY REGARDING THE ESA S2P AND EU-SST PROGRAMS

NELIOTA will continue the scheduled operations until June 2023 as a collaborator of the ESA/CARMEN project. However, a possible extension of NELIOTA is under discussion with ESA's stakeholders. Moreover, in case of additional funding, except for the operational costs, the procurement of new more sensitive and preferably faster read-out sCMOS cameras will be examined. The latter will allow to reduce the exposures (i.e. to increase the frame rates from 30 fps to 40 fps), thus, to obtain a better coverage of the light curves of the multiframe flashes. The NELIOTA team plans to publish in 2023 the update of the statistics regarding the physical parameters and the appearance frequencies of the meteoroids based on the detected lunar impact flashes. Moreover, an extensive study on the meteoroid streams' projectiles is under discussion in order to try to constrain more the luminous efficiency ranges, thus, to constrain more the physical parameters of the meteoroids.

Regarding the data collection from other observers using the *Flash Detection Software* (FDS; see section 2.2), a data archive center will be certainly needed. For the latter, the establishment of a dedicated team is needed in order: a) to check the detections based on heterogeneous data (i.e. different cameras-telescopes set-ups), b) to validate the flashes that were observed simultaneously from different locations, and c) to establish and maintain a database and publish the results (probably via a website).

The 1.2 m Kryoneri telescope is part of the Europlanet Telescope Network (EPN-TN), and we anticipate amateur or professional observers to make use of the instrument for their projects. Obviously, the high-cadence twin camera system can be used for a lot of projects, especially for asteroid/NEOs photometry and occultations (e.g. [23]) but also for astrometry.

The Kryoneri team has been collaborating with various international teams regarding the tracking (i.e. astrometry) and photometry of asteroids and NEOs. Between 2019-2021, Kryoneri was funded by ESA for observations of NEO occultations. We expect that this kind of collaborations will increase in the near future and the 1.2 m Kryoneri telescope will play a key role for the S2P program of ESA. Fig. 17 shows the Didymos-Dimorphos binary asteroid system approximately one month after the impact of the NASA-DART mission.

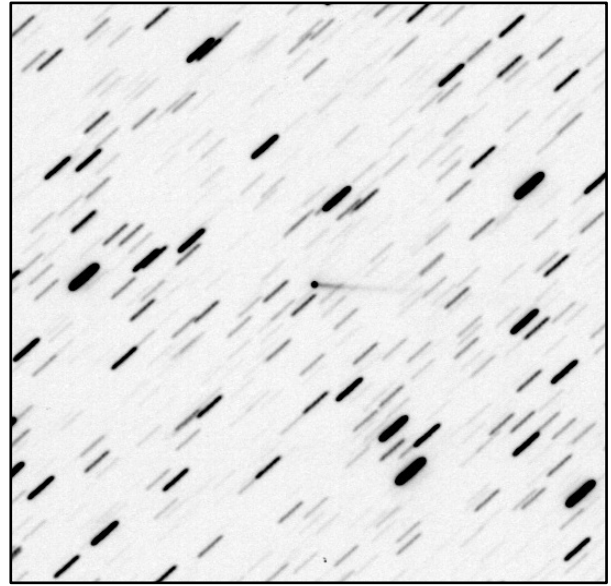

*Figure 17. The Didymos-Dimorphos asteroid binary system as observed from Kryoneri telescope and the Zyla I-camera on 20-Oct-2022. The system is the point source at the center and its debris tail, formed due to the NASA-DART impact one month earlier, has a direction from the center to the right part of the image. The trails are field stars. The image is a stack of ten frames.*

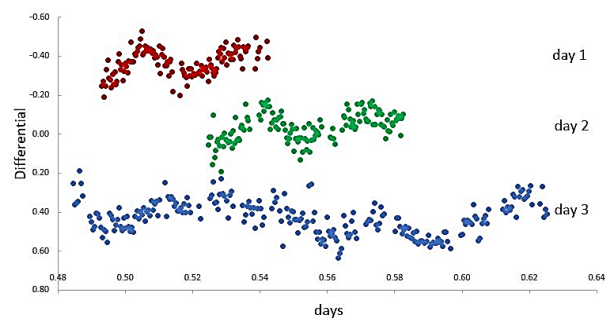

*Figure 18. Photometry of the Didymos-Dimorphos asteroid binary system for three successive nights as resulted from Kryoneri telescope observations. The periodic behaviours of the points correspond to the rotation of the system's components.*

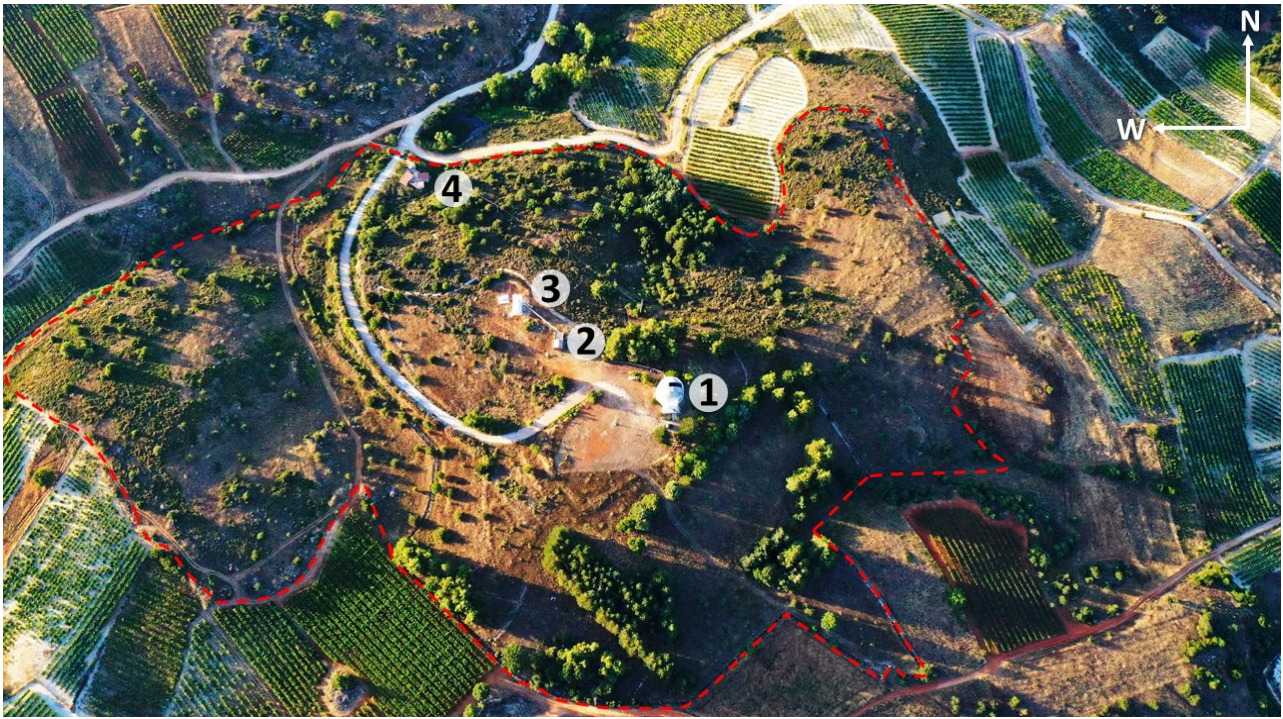

*Figure 19. Aerial view of the Kryoneri Observatory area that covers approximately 93,000 m² (within the red dashed lines). Explanation of indices: 1) main observatory building (1.2 m telescope), 2) power generator, 3) auxiliary building, 4) old guesthouse*

The photometry of the system for three consecutive nights is plotted in Fig. 18. Our team plans to continue the monitoring of this system aiming to determine its orbital period change rate as well as to initiate further collaborations for observations of the target asteroids of the ESA Hera mission.

Moreover, in 2021, the 1.2 m Kryoneri telescope was part of an international collaboration, led by ESA members, aiming to the astrometric position monitoring of the ESA-BepiColombo mission in order to simulate the spacecraft's trajectory as a potential NEO threat [24].

In early 2023, NOA plans to procure and install two fireball cameras in order to participate in the European AllSky7 fireball network[8]. The one camera will be installed at the Kryoneri observatory area, while the second one probably at the headquarters of NOA/IAASARS at Penteli, Athens, Greece. Depending on future funding, NOA aims to also join other fireball networks such as FRIPON[9] and the Global Meteor Network[10]. All these aforementioned networks have no sensors in the region of Southeast Europe. Therefore, it is anticipated that our participation in these networks, taking into account the high amount of cloudless nights per year, will significantly increase the detections of such transient phenomena allowing to enrich our knowledge on this topic.

Regarding the SST activities, NOA plans to completely upgrade the Kryoneri Observatory premises. Except for the operations of the 1.2 m telescope, new instruments and infrastructure are planned to be installed at Kryoneri observatory in the following 2-4 years. An example of these plans concerns the procurement of a new telescope with diameter between 60-70 cm with a fast tracking mount that will allow for observations of LEO satellites as well. This telescope is planned to be fully robotic and will be equipped with two cameras, particularly a sCMOS for the satellite tracking and a CCD for other astronomy related projects (e.g. NEOs).

In addition, NOA also plans to build a public outreach center at Kryoneri Observatory in order to disseminate the scientific results to the public. The building works are anticipated to begin in 2023. Fig. 19 shows a top view of the Kryoneri Observatory area indicating the current facilities.

## 5    ACKNOWLEDGEMENTS

A.L. acknowledges financial support by the European Space Agency (ESA) under the Consolidating Activities Regarding Moon, Earth and NEOs (CARMEN) project, NOA/SRFA no. 1084. We thank the NOA-BEYOND team and particularly Ms. Stella Girtsou

---

[8] https://allsky7.net/
[9] https://www.fripon.org/
[10] https://globalmeteornetwork.org/

---

(NOA/IAASARS) for her help in the orbits predictions and the preparation of the TDM files for the calibration tests of 1.2 m Kryoneri telescope. We are grateful to Prof. Kleomenis Tsiganis (Aristotle University of Thessaloniki, Greece) for his assistance in the observations setup with the Kryoneri telescope software for satellite tracking and the astrometric data analysis. The 1.2 m Kryoneri telescope is operated by the Institute for Astronomy, Astrophysics, Space Applications and Remote Sensing of the National Observatory of Athens. The *Flash Detection software* was developed by S. Achlatis, G. Christofidi, and I. Chatzi under the ESA Contract Nr. 4000135574/21/NL/IB/gg.